\documentclass[fleqn,usenatbib]{mnras}
\usepackage[latin9]{inputenc,stmaryrd}
\usepackage{graphicx,environ}
\usepackage{amsmath,mathtools}
\usepackage{amssymb}
\usepackage{verbatim}
\usepackage{ulem,cancel}
\usepackage{newtxtext,newtxmath}
\usepackage{ae,aecompl}
\usepackage{yfonts}
\usepackage{float}
\usepackage{epstopdf}
\usepackage{graphicx}
\usepackage{pdfpages}

\def\degree{\ifmmode {^\circ}\else {$^\circ$}\fi}
\def\mum{\ifmmode {\rm \mu {\rm m}}\else $\rm \mu {\rm m}$\fi}
\def\arcsec{\ifmmode ^{\prime \prime}\else $^{\prime \prime}$\fi}

\def\inch{\ifmmode ^{\prime \prime}\else $^{\prime \prime}$\fi}
\def\arcmin{\ifmmode ^{\prime}\else $^{\prime}$\fi}

\def\mjup{\ifmmode {\rm M_J}\else $\rm M_J$\fi}
\def\rjup{\ifmmode {\rm R_J}\else $\rm R_J$\fi}
\def\mearth{\ifmmode {\rm M_{\oplus}}\else $\rm M_{\oplus}$\fi}
\def\rearth{\ifmmode {\rm R_{\oplus}}\else $\rm R_{\oplus}$\fi}
\def\lsun{\ifmmode {\rm L_{\odot}}\else $\rm L_{\odot}$\fi}
\def\msun{\ifmmode {\rm M_{\odot}}\else $\rm M_{\odot}$\fi}
\def\mjupyr{\ifmmode {\rm M_J~yr^{-1}}\else $\rm M_J~yr^{-1}$\fi}
\def\msunyr{\ifmmode {\rm M_{\odot}~yr^{-1}}\else $\rm M_{\odot}~yr^{-1}$\fi}
\def\kms{\ifmmode {\rm km~s^{-1}}\else $\rm km~s^{-1}$\fi}

\def\gcm3{g~cm$^{-3}$}

\NewEnviron{myequation}{
\begin{equation}
\scalebox{0.85}{$\BODY$}
\end{equation}}

\title[Planet migration in wind-fed accretion disks in binaries]{Planet migration in wind-fed accretion disks in binaries}

\author[Kulikova, Popov, Zhuravlev]{
Olga Kulikova$^{1}$,
Sergei B. Popov$^{1,2}$, Viacheslav V. Zhuravlev$^{1}$\\
$^1$ Sternberg Astronomical Institute, Lomonosov Moscow State University, \\ Universitetskij prospekt 13, Moscow, 119234 Russia\\
$^2$ National Research University ``Higher School of Economics'', Department of Physics\\
Myasnitskaya str. 20, Moscow 101000, Russia
}

\date{Accepted XXX. Received YYY; in original form ZZZ}

\pubyear{201*}

\begin{document}
\maketitle
\begin{abstract}
Planet migration originally refers to protoplanetary disks, which are more massive and dense than typical accretion disks in binary systems.
We study planet migration in an accretion disk in a binary system consisting of a solar-like  star hosting a planet and a red giant donor star. The accretion disk is fed by a stellar wind.
We use the $\alpha$-disk model and consider that the stellar wind is time-dependent. Assuming the disk is quasi-stationary we calculate its temperature and surface density profiles. In addition to the standard disk model, when matter is captured by the disk at its outer edge, we study the situation when the stellar wind delivers matter on the whole disc surface inside the accretion radius with the rate depending on distance from the central star. Implying that a planet experiences classical type I/II migration we calculate migration time for a planet on a circular orbit coplanar with the disk. Potentially, rapid inward planet migration can  result in a planet-star merger which can be accompanied by an optical or/and UV/X-ray transient. 
We calculate timescale of migration for different parameters of planets and binaries. Our results demonstrate that planets can fall on their host stars within the lifetime of the late-type donor for realistic sets of parameters. 

\end{abstract}
\begin{keywords}
 binaries: close -- accretion, accretion discs -- planet-disc interactions
\end{keywords}

\section{Introduction}
Modern observations demonstrate a plethora of different exoplanets and extrasolar planetary systems. The number of confirmed exoplanets has exceeded $\sim 3500$\footnote{Open Exoplanet Catalogue \url{http://www.openexoplanetcatalogue.com/} }  at the time of writing this paper.
Approximately two thirds of stars (the exact fraction depends on masses) are members of binary or multiple systems \citep[e.g.][]{ragh+10}. They can also host planetary systems which are divided in two types: p-type (when a planet orbits the whole binary) and s-type (when a planet orbits one member of the system).
Now $>100$ of known exoplanets are found in binary (or multiple) systems \citep{schw+16}. 
  
Multiple, and in particular binary, systems hosting planets present a specific laboratory for testing planet formation and evolution scenarios. The presence of a companion star would affect the process of planet formation and evolution. In this paper we consider a planet on an s-type orbit around a secondary (initially less massive) solar-like star. A planet can survive dynamical evolution of the binary up to a specific moment when an accretion disk is formed around its host star due to a stellar wind from the primary component. 

Gravitational perturbations due to presence of a stellar companion might be crucial for planet formation either via core accretion or by protoplanetary disk instability, especially in the case of close binaries with separations less than $\sim (20-100)$ AU. Nevertheless, survival  of planets in some close binaries with separations of tens of AU suggests that these perturbations are not always severe enough to preclude planet formation \citep[see e.g.][]{exoplanet2,gliese,exoplanet1}. 


The presence of a stellar companion results in truncation of a circumstellar
protoplanetary disk \citep{planets_in_binaries}, hence limiting planet formation and reducing planet occurrence by a factor of two or even larger for very compact binary systems \citep{wang14}. 

In this study we are interested in systems where the primary (initially more massive companion of a binary) is an evolved donor star that experiences a mass loss via intense low-velocity stellar wind resulting in 
an accretion disk formation around the secondary. 
Mass transfer from an evolved donor to its companion star is a typical result of binary evolution. Formation of a disk around the acceptor yields a significant impact on the evolution of the planetary system around it. The lifetime of an accretion disk depends mainly on the duration of the red giant phase of the primary and hence on its mass. For primaries considered in this study, lifetime of accretion disks is about 10 Myr and is comparable  with a typical protoplanetary disk lifetime which is $\sim$2-3 Myr \citep{protopl_disk_lifetime}. Thus, such accretion disk is a relatively long-living structure significantly influencing  evolution of embedded planets. The newly formed disk around a Main sequence (MS) companion might interact with a pre-existing planet, resulting in its growth and migration.

Wind-fed discs in binaries have been studied for many decades with different techniques (analytical and numerical models with variety of assumptions). \cite{perken13} studied evolution of a non-stationary disk and obtained that disks in binaries with separations less than 100 AU, and with masses of a donor and an accretor equal to 3 M$_{\odot}$ and 1 M$_{\odot}$, respectively, reach steady state within 1-2 Myr. 
 Under realistic assumptions it can be shown that the viscous timescale corresponding to an active disk supplied by a stellar wind at a rate in the range 10$^{-5}-10^{-8}$ M$_\odot$~yr$^{-1}$ does not exceed the wind variation timescale in the case of a red giant donor with $M\approx (3-5)$ M$_\odot$. Thus, in this paper we employ quasi-stationary disk models.
We use two disk models. The first model is the standard accretion disk driven by effective viscosity described by \cite{shakura_sunyaev} (Sec. \ref{SD}). The second model is also described using the $\alpha$-disk approach, but unlike the first model it incorporates wind matter settling, stellar irradiation, and dust evaporation (Sec. \ref{DMS}).

Planet-disk interactions in binaries are usually studied in terms of a planet embedded in a circumbinary or circumstellar protoplanetary disks (see \citealt{kley_survey} and references therein). In this work we consider a planet experiencing type I and type II migration in disks of  different origin and with distinct properties rather than protoplanetary disks.  We study a circumstellar disk formed by wind matter captured from an evolved companion star. Matter is continuously added to the disk, and then is accreted onto the MS star hosting the planet. The planet evolves in this disk and, finally, may coalesce with the star.  

For type I migration we use equations given in \cite{tanaka} which correspond to a quite rapid orbital evolution. If inward planet migration is effective enough then it results in a planet-star direct impact, or a planet can be tidally destroyed by the host star, or stable accretion from the planet to the star can be initiated, depending on the ratio between planetary and stellar average density (see e.g. \citealt{metzger}). The duration of the planet migration strongly depends on the surface density and the scale height of the disk. Expected migration time in the accretion disk in an evolved binary is larger than in a typical protoplanetary disk since the former has lower density than the latter one. In this study we calculate the timescale of planet migration for various masses of stellar components, different binary separations, and planetary masses in order to compare it with the accretion disk lifetime defined by the lifetime of a red giant (RG) donor star. 

In the following section we describe two models of wind-fed accretion disks which we use in this study. Next, in Sec. 3 equations of the planetary migration are presented. Then in Sec. 4 our results are given. They are discussed in Sec. 5. Finally, we summarize our conclusions in the final section.

\section{Wind-fed accretion disk} \label{Wind-fed accretion disk}

In this section we present our approach to calculate accretion disk properties.

\subsection{Accretion rate}

Defining mass loss rate of the donor as $\dot{M}_\mathrm{w}$, we suppose that the disk viscous timescale is less then the timescale of wind variations: $t_\mathrm{vis}\ll t_{\dot{\mathrm{M}}_\mathrm{w}}$, here $t_{\dot{\mathrm{M}}_\mathrm{w}}=\dot{M}_\mathrm{w}/\ddot{M}_\mathrm{w}$. We build a quasi-stationary model of a geometrically thin $H\ll r$ accretion disk, $r$ is a radial distance from the star, and $H$ is the vertical scale height at this $r$.
The disk matter is moving in Keplerian orbits around a MS star of mass $M_\mathrm{2}$ with orbital angular velocity $\Omega=\sqrt{GM_{2}/r^{3}}$.
We use $\alpha$-viscosity approach that was first described by \cite{shakura}. The viscosity is parameterized as:
\begin{equation}
\label{nu}
\nu=\alpha c_\mathrm{s} H,
\end{equation}
with viscosity parameter $\alpha$, where $c_\mathrm{s}$ is the sound speed in the disk midplane related to $H$ as: 
\begin{equation}
\label{vert_eq}
H = c_\mathrm{s}\Omega^{-1},
\end{equation}
according to vertical hydrostatic equilibrium. The sound speed is: 
\begin{equation}
\label{c_s}
c_\mathrm{s}=\sqrt{ \frac{\gamma kT_\mathrm{c}}{\mu m_\mathrm{H}} },
\end{equation}
where $\gamma$ is the specific heat ratio, $\mu$ is the mean molecular weight, $k$ is the Boltzmann constant, $T_\mathrm{c}$ is the disk midplane temperature, and $m_\mathrm{H}$ is the hydrogen atom mass. We follow \cite{soker_rappaport} to reveal conditions of disk formation from wind in a binary. Assuming $M_{2}\sim$ M$_{\odot}$ and radius of the MS star $R_{2}\sim$ R$_{\odot}$, wind accretion disks are formed in binaries with $a\lesssim$ 10--100 AU for M$_{1}\approx$ 1--10 \msun\ and relative velocity of wind and accretor $v_\mathrm{rel}\approx~5-20$~\kms. The relative velocity of the wind here is defined as $v_\mathrm{rel}=\sqrt{v_\mathrm{w}^2+v_\mathrm{s}^2}$, $v_\mathrm{w}$ is the wind velocity  at location of the secondary star and
$v_\mathrm{s}=\sqrt{GM_1^2/[M_1+M_2]/a}$ is the orbital velocity of the secondary (implying that the origin is set to the center of mass of the system).


Typical mass loss rates of RG tars are $\dot{M}_{w}\sim10^{-9}-10^{-6}$ \msunyr\ \citep{ram+06}.
The mass loss rate of a donor star in our calculations is obtained using re-derived classical Reimers law \citep{dotMwind}:
\begin{equation}
\dot{M}_\mathrm{w}=\eta  (L_1/L_\odot) (R_1/R_\odot) (M_1/M_\odot)^{-1} \,\left(\dfrac{T_{1}}{4000~\mathrm{K}}\right)^{3.5}\,\left(1+\dfrac{g_{\odot}}{4300~g_1}\right),\label{eq:dotMw}
\end{equation}
here $\eta=8\times 10^{-14}$ M$_{\odot}~\mathrm{yr}^{-1}$, g$_{\odot}$ --- solar surface gravitational acceleration, $L_1$, $R_1$, and $M_1$ 
refer to luminosity, radius, and mass of the donor star. Mass loss rate depends on effective temperature $T_1$ and surface gravitational acceleration $g_1$ of the donor star. In order to obtain $L_1$, $R_1$, $M_1$, and $T_1$ we use stellar evolutionary tracks calculated by the Padova group \citep{padova}\footnote{Padova stellar evolutionary tracks are available on-line at \url{http://pleiadi.pd.astro.it/} }. 

To estimate the accretion rate by the secondary we use the Bondi-Hoyle rate for an isotropic wind from the donor:
\begin{equation}
\dfrac{\dot{M}_\mathrm{acc}^\mathrm{tot}}{\dot M_\mathrm{w}}=\left(\dfrac{r_\mathrm{a}}{2a}\right)^{2},\label{eq:mdot_acc}
\end{equation}
where the Bondi accretion radius $r_\mathrm{a}$
 is defined as: 
\begin{equation}
r_\mathrm{a}=\dfrac{2GM_{2}}{v_\mathrm{rel}^{2}+c_\mathrm{w}^{2}}, \label{eq:R_a}
\end{equation}
here c$_{\mathrm w}$ is the sound speed in wind.

 \subsection{Standard disk} \label{SD} 
The first model that we use will be referred as the {\it standard disk} (SD).
In the SD model matter enters the disk formally at infinite distance.
The continuity equation yields the constant mass accretion rate across the disk:

\begin{equation}
\dot{M} = \dot{M}^\mathrm{tot}_\mathrm{acc}.
\label{cont_eq}
\end{equation}
 
Together with eq. (\ref{cont_eq}) the angular momentum conservation law 
yields the following expression  \citep{shakura_sunyaev}:
\begin{equation}
\nu\Sigma=\frac{\dot{M}^\mathrm{tot}_\mathrm{acc}}{3\pi} \, f, \label{nu_sigma_SD}
\end{equation}
where $\Sigma$ is the surface density of the disk and
\begin{equation}
\label{f_SD}
f = 1-\left(\dfrac{r_\mathrm{in}}{r}\right)^{1/2} 
\end{equation}
is a usual additional factor originating from the zero-torque boundary condition, which tends to unity far from the inner boundary of the disk.

 Equations (\ref{nu}), (\ref{vert_eq}), (\ref{c_s}) and (\ref{nu_sigma_SD}) enable to express the surface density and the vertical scale height in terms of $T_\mathrm{c}$ as follows: 
\begin{equation}
\Sigma=\dfrac{m_\mathrm{H} \mu }{3\pi\gamma k_\mathrm{B}} \Omega f \dot{M}_\mathrm{acc}^\mathrm{tot} \alpha^{-1}T_\mathrm{c}^{-1},
\label{Sigma_T}
\end{equation}
\begin{equation}
H=\left(\dfrac{\gamma k_\mathrm{B} T_\mathrm{c}}{\mu m_\mathrm{H}\Omega^2}\right)^{1/2}.
\label{H_T}
\end{equation}

The midplane disk temperature in a SD is determined by $T_\mathrm{c}=T_\mathrm{v}$, where the ``viscous'' temperature $T_\mathrm{v}$ is estimated for {\it optically thick} disk as \citep{rud+86}: 
\begin{equation}
T_\mathrm{v}=\left(\frac{27\kappa\nu \Omega^{2}\Sigma^{2}}{64\sigma}\right)^{1/4} ,
\label{eq: t-visc1}\end{equation}
 where $\sigma$ is the Stefan-Boltzmann constant.
Here, the  midplane optical depth is estimated as $\tau=\kappa\Sigma/2$ with 
the opacity assumed to be constant $\kappa=\kappa_{0}$, where $\kappa_0=2$ cm$^{2}~$g$^{-1}$ is a typical 
value for cool disk material of solar metallicity filled with dust grains.\footnote{In this study we discuss only planet migration within the snow line zone, i.e. we neglect complications of opacity in cold regions of the disk.}

Equation (\ref{eq: t-visc1}) 
allows us to obtain the following disk profiles explicitly: 
\begin{equation}
\Sigma=4.6~\alpha^{-4/5} m^{1/5} \dot{m}^{3/5}f^{3/5}\tilde{r}^{-3/5} 
\mathrm{~g~cm^{-2}},\end{equation}
\begin{equation}
H=8.5~\times 10^{11}~\alpha^{-1/10} m^{-7/20} \dot{m}^{1/5} f^{1/5}\tilde{r}^{21/20} \mathrm{~cm},
\end{equation}
where $m=M_2/M_\mathrm{\odot}$, $\dot{m}=\dot{M}_\mathrm{acc}^\mathrm{tot}/[10^{-10}~M_\mathrm{\odot}~\mathrm{yr}^{-1}]$ 
and $\tilde{r}=r/$AU.

 \subsection{Accretion disk with matter settling} \label{DMS}
 
The second approach that we apply in this study is based on a concept of a {\it disk with matter settling} (DMS).

In this case the wind matter settles onto the disk surface at all radii within the Bondi accretion radius $r_\mathrm{a}$.
Following \cite{perken13} we define the source of matter in a disk as:
 \begin{equation}
\dot{\Sigma}_\mathrm{ext}(r)=
\left\{  
\begin{array}{ll}
\dfrac{\dot{M}_\mathrm{acc}^\mathrm{tot} }{2\pi r r_\mathrm{a}}, & r<r_\mathrm{a}\\
~0.& r\geq r_\mathrm{a}
\end{array} \right.\label{eq:dotSigma(R)}\end{equation}





Then, the continuity equation in the disk takes the following form:
\begin{equation}
\dfrac{1}{r}\dfrac{\partial}{\partial r}\dot{M}(r)=-\dot{\Sigma}_\mathrm{ext}.\label{eq:continuity}
\end{equation}

The mass flow rate within the disk is defined as:
\begin{equation}
\dot{M}(r)=-2\pi r\Sigma v_\mathrm{r}, \label{dotM}
\end{equation}
where $v_\mathrm{r}$ is the radial velocity in the disk.

Angular momentum conservation law in the disk can be written as:
\begin{equation}
\dfrac{1}{r}\dfrac{\partial}{\partial r}\left(r^3\Omega
\Sigma v_r\right)=\dfrac{1}{2 \pi r}\dfrac{\partial \mathcal{G}}{\partial r} + \Omega r^2\dot{\Sigma}_\mathrm{ext}\label{ang_mom_conserv},
\end{equation}
where we define the viscous torque:
$$
\mathcal{G} = 2\pi \nu \Sigma r^3 d \Omega / dr.
$$
Using the continuity equation and the zero torque condition at the inner disk boundary $\nu \Sigma \left(r_\mathrm{in}\right)=0$
we obtain:
\begin{equation}
\nu \Sigma =\dfrac{1}{3\pi r^2\Omega}\left(\dot{M}\Omega r^2\bigg|^r_{r_\mathrm{in}}+\int^r_{r_\mathrm{in}}2\pi \Omega r^3 \dot{\Sigma}_\mathrm{ext}dr \right)\label{nu_sigma}.
\end{equation}
It can be shown that  eq. (\ref{nu_sigma}) reduces to the expression (\ref{nu_sigma_SD}) if we set $\dot{\Sigma}_\mathrm{ext}=0$.
If one uses the explicit expression for the mass source then  eq. (\ref{nu_sigma}) is reduced to the usual expression for standard disk,  see eq. (\ref{nu_sigma_SD}), with $f$ given as: 

 \begin{equation}
 \resizebox{0.908\hsize}{!}{$
 f=
 \begin{cases} 
\dfrac{r_\mathrm{a}+r_\mathrm{in}-r}{r_\mathrm{a}}-\left(\dfrac{r_\mathrm{in}}{r}\right)^{1/2}+\dfrac{2}{3}\dfrac{r}{r_\mathrm{a}}\left[1-\left(\dfrac{r_\mathrm{in}}{r}\right)^{3/2}\right],& r<r_\mathrm{a}\\
\dfrac{r_\mathrm{in}}{r_\mathrm{a}}-\left(\dfrac{r_\mathrm{in}}{r}\right)^{1/2}+\dfrac{2}{3}\left(\dfrac{r_\mathrm{a}}{r}\right)^{1/2}\left[1-\left(\dfrac{r_\mathrm{in}}{r_\mathrm{a}}\right)^{3/2}\right], & r\geq r_\mathrm{a} 
\end{cases}  \label{f}  .
$}
\end{equation}

Note that away from both the inner boundary (i.e., at $r\gg r_\mathrm{in}$) and the Bondi accretion radius (i.e., at $r\ll r_\mathrm{a}$) the expression (\ref{f}) simplifies to
\begin{equation} 
f=1+\dfrac{1}{3}\dfrac{r}{r_\mathrm{a}}.\label{f_r_in<<r<<r_a}
\end{equation}

In the DMS model we take into account an additional heating of the disk, which occurs due to radiation from the host star and dissipation of the kinetic energy of wind material falling onto the disk surface.
In order to calculate the midplane disk temperature in this case, we use the following equation:
\begin{equation}
T_\mathrm{c}^{4}=T_\mathrm{v}^{4}+T_\mathrm{irr}^{4}+T_\mathrm{fall}^{4},
\label{eq: tdisk}\end{equation}
where $T_\mathrm{v}$ is defined as in the case of the SD model by eq. (\ref{eq: t-visc1}), whereas
$T_\mathrm{irr}$ and $T_\mathrm{fall}$ are, respectively, the irradiation temperature due to heating from the host star, and the temperature due to disk heating by the wind material falling onto the disc surface. We write eq. (\ref{eq: tdisk}) 
following \citet{Calvet91} (see their Section 2). These authors provide a justification for the superposition treatment of solutions for transfer equation in the case when additional external heat is fully radiated back from the disk photosphere.

With regards to the heating by the falling wind material, see \cite{perken13}, the thermal energy input into an annulus of a unit area at radius $r$ is:
\begin{equation}
\sigma T_\mathrm{fall}^4=\left\{
\begin{array}{ll}
\dfrac{GM_2 \dot{M}_\mathrm{acc}^\mathrm{tot}}{2  \pi r^2 r_\mathrm{a}}, & r<r_\mathrm{a}\\
0, & r\geq r_\mathrm{a}
\end{array} \right.,\label{eq:F_accr}\end{equation}
 which defines $T_{\mathrm{fall}}$.


Next, in vertically isothermal disk the irradiation temperature $T_\mathrm{irr}$ 
is a function of the effective temperature of the star, $T_2$, and the star radius $R_2$ \citep{irr_temp}:
\begin{equation}
\dfrac{T_\mathrm{irr}^4}{T_2^4}=\dfrac{2}{3\pi}\left(\dfrac{R_2}{r}\right)^3+\dfrac{H}{2r}\left(\dfrac{R_2}{r}\right)^2\left(\dfrac{\partial{\ln H}}{\partial{\ln r}}-1\right).\label{eq: Tirr}
\end{equation} 


Let us impose $\partial{\ln H}/\partial{\ln r}=9/7$ following \cite{irr_temp}. 
Using eq. (\ref{vert_eq}) and 
eq. (\ref{c_s}),
we obtain $T_\mathrm{irr}$ as a function of $T_\mathrm{c}$.

We solve equation (\ref{eq: tdisk}) numerically using Newton-Raphson technique following the approach by \cite{perken13}. 

Additionally, for the DMS model we take into account that in the region near the secondary component the disk temperature increases enough to vaporize dust implying a change in opacity \citep{cha09}. The opacity of the disk matter is determined as:
\begin{equation}
\kappa=\kappa_\mathrm{0}\left(\dfrac{T_\mathrm{c}}{T_\mathrm{e}}\right)^{m_\mathrm{k}}
\label{kappa},
\end{equation}
where $T_\mathrm{e}$=1380K is the dust evaporation temperature and $\kappa_\mathrm{0}$ is defined in the previous Section. 
The exponent in the eq. (\ref{kappa}) is defined as $m_\mathrm{k}=-14$ when $T_\mathrm{c}>T_\mathrm{e}$ \citep{rud+91} and $m_\mathrm{k}=0$ for $T_\mathrm{c}<T_\mathrm{e}$. Every time the former inequality is satisfied, we check 
that $\tau$ does not fall below unity. Even though eq. (\ref{kappa}) introduces the drastic decrease of opacity, it was checked that the inner parts of DMS remain optically thick in the range of accretion rates considered here.

The surface density and vertical scale height as functions of $T_\mathrm{c}$ are still given by eqs. (\ref{Sigma_T}) and (\ref{H_T}) with the 
factor $f$ defined by eq.~(\ref{f}).


The equations above do not depend on the value of the outer disk radius. Formally, we consider that it is limited by the Roche limit of the secondary $r_\mathrm{out}$ \citep{egg83}:
\begin{equation}
\dfrac{r_\mathrm{out}}{a}=\frac{0.49q^{2/3}}{0.6q^{2/3}+\ln(1+q^{1/3})}.~\label{eq:R_roche}\end{equation}
We consider planets at the distance from the host star $r_\mathrm{p}<r_\mathrm{out}$.
It is assumed that the semi-major axis of the binary, $a$, is constant over time. 

Young stars have strong enough magnetic fields to truncate the disk quite far from 
the star surface.
In this work we consider an evolved MS star with a magnetic field which is generally negligible for the accretion dynamics. At the same time, the width of the boundary layer, which forms between the star surface and the disk, $\Delta r_\mathrm{bl}/r\sim(H/r)^{2}$, 
\citep{armitage}. Thus, $\Delta r_\mathrm{bl}\ll R_{2}$, so we set the inner boundary of the disk 
be equal to the radius of the secondary: 
\begin{equation}
r_\mathrm{in}=R_2.
\label{eq:R_in}
\end{equation}

\section{Planet migration in gaseous disk} \label{Planet migration in gaseous disk}

Accretion disk interacts with an embedded planet via gravitational torques 
which cause planet migration. 
The lifetime of an accretion disk is constrained by the duration of the RG phase of the mass losing star $\gtrsim  10^7$ yrs for $M_1\lesssim 5 $ M$_\odot$, which is longer than the lifetime of a protoplanetary disk and allows  migration of planets to become even more considerable than at the birth of a planetary system.




The detailed physics of the planet-disk gravitational interaction is a complicated 
topic (see e.g. the review by \citealt{Paardekooper2018}). Planetary mass is an important parameter that defines the regime of  migration. Type I migration occurs in the case of a low-mass planet whose interaction with disk matter generates weak density waves which in general carry away the planet orbital angular momentum in the linear regime.  
In the case of a high-mass planet 
its gravitational torques are strong enough to create 
an annular gap around the planet. This corresponds to type II migration which occurs at the rate of viscous transfer of angular momentum in a disk.


\subsection{Type I migration}

In order to calculate the total torque acting on a planet of a  mass $M_\mathrm{p}$ at a distance $r_\mathrm{p}$ from the host star in the disk with surface density $\Sigma(r)\sim r^\mathrm{-\bar{\alpha}}$  and thickness $H(r)$ we use the result of \citet{tanaka}:
\begin{equation} 
\dfrac{dJ_\mathrm{p}}{dt}=-\left(1.36+0.54\bar{\alpha}\right)\left(\dfrac{M_\mathrm{p}}{M_2}\right)^2\left(\dfrac{H}{r_\mathrm{p}}\right)^{-2}\Sigma r_\mathrm{p}^4 \Omega_\mathrm{p}^2,\label{eq: type_I}
\end{equation}
where $\Omega_\mathrm{p}(r_\mathrm{p}$) is the Keplerian frequency at $r_\mathrm{p}$
and  $J_\mathrm{p}=M_\mathrm{p}\Omega_\mathrm{p} r_\mathrm{p}^2$ is the angular momentum of a planet in a Keplerian orbit. Eq. (\ref{eq: type_I}) is obtained for the basic model case of three-dimensional isothermal disk. 
The corresponding rate of planet migration is:
\begin{equation}
\dfrac{dr_\mathrm{p}}{dt}=-\left(2.72+1.08\bar{\alpha}\right) \dfrac{M_\mathrm{p}}{M_2^2} \left(\dfrac{H}{r_\mathrm{p}}\right)^{-2}\Sigma r_\mathrm{p}^3 \Omega_\mathrm{p}.\label{eq: type_I_speed}
\end{equation}

Eq. (\ref{eq: type_I_speed}) is valid for a planet embedded into 
a disk around a single star. However,
 tidal effects of a stellar companion on the disk material may be crucial for a subtle physics of planet-disk interaction. 
Thus, one should compare disturbing forces acting on a disk from the planet and the companion. 
The condition that gravitational interaction between the planet and disk material close to the planet is stronger than the tidal action of the companion reads: 
\begin{equation}
\label{tidal}
GM_\mathrm{p}/\bar{r}^2 > GM_1 r_\mathrm{p}/a^3,
\end{equation} 
where $\bar{r}=2H/3$ is the approximate location of the strongest resonances that exist in an annulus around the planet (\citealt{ward98,Artymowicz1993}). Eq. (\ref{tidal}) enables us to put the lower limit on the planet-to-primary-star mass ratio $q^\prime \equiv M_\mathrm{p}/M_2$. 
This lower limit on $q^\prime$ is the following:
\begin{equation}
q^\prime_\mathrm{min}=\dfrac{4}{9} \frac{M_1}{M_2} \left(\dfrac{H}{r_\mathrm{p}}\right)^2\left(\dfrac{r_\mathrm{p}}{a}\right)^3  \label{eq: q_min}.
\end{equation}

The value of minimum planet mass defined by eq. (\ref{eq: q_min}) at $r_{\mathrm{p}}\sim$ 1 AU does not exceed $\sim 0.005$  M$_\mathrm{Jup}$ for systems with the 5 M$_\odot$ donor and rapidly vanishes as the planet approaches the host star. 

Additionally, the dynamical interaction of stellar companion with the planet imposes the maximum semi-major axis for stable orbits of  s-type planets \citep{dyn_stability}:
\begin{equation}
\dfrac{r_\mathrm{p}^\mathrm{max}}{a}=0.464-0.38~\dfrac{M_1}{M_1+M_2}.
\end{equation}

We stop planet migration at $r_\mathrm{p}^\mathrm{fin}$ which is defined as:
\begin{equation}
r_\mathrm{p}^\mathrm{fin}=\left(R_\mathrm{in}^{13/2}+29.25\sqrt{GM_2}R_{2}^5\dfrac{M_p}{Q}t_\mathrm{RG}\right)^{2/13} \label{r_fin}
\end{equation}
where the tidal dissipation parameter Q is set to $10^{5.5}$, while $t_\mathrm{RG}$ is duration of the red giant phase which defines the lifetime of an accretion disk.
The expression (\ref{r_fin}) is obtained by integration of expression (2) in \cite{Jackson} (considering e=0).

\subsection{Type II migration}
The planet transfers angular momentum to the outer disk ($r\gtrsim r_\mathrm{p}$) and gains it from the inner disk ($r\lesssim r_\mathrm{p}$). If the absolute value of torque exerted by the planet on disk is larger than the viscous torque then an annular gap forms.
\cite{baruteau} obtained the critical planet-to-star mass ratio $q^\prime$ for gap opening:
\begin{equation}
q^\prime_\mathrm{crit}=\dfrac{100}{\mathcal{R} }\left[\left(X+1\right)^{1/3}-\left(X-1\right)^{1/3}\right]^{-3},\label{eq: q_crit}
\end{equation}
where $X=\sqrt{1+3\mathcal{R}  H^3/(800 \, r_\mathrm{p}^3)}$ and $\mathcal{R} =r_\mathrm{p}^2\Omega_\mathrm{p}/\nu$ is the Reynolds number corresponding to effective viscosity in a disc.
In order to analyze type II migration we consider planets with $q^\prime>q_\mathrm{crit}^\prime$. 

In this regime a planet migrates inwards together with the local disk material 
at the rate equal to the radial velocity of gas in an unperturbed disk regardless of the mass of the planet:  
\begin{equation}
\dfrac{dr_\mathrm{p}}{dt}=-\dfrac{3}{2}\alpha\left(\dfrac{H}{r_\mathrm{p}}\right)^2 f^{-1} v_\mathrm{K},\label{eq: type_II}
\end{equation}
where $v_\mathrm{K}$ is Keplerian velocity in the disk at the distance $r_\mathrm{p}$, while $f$ is defined 
by eq. (\ref{f_SD}) and eq. (\ref{f}) for, respectively, SD and DMS.

\section{Results}
To obtain the time of planet migration in wind-fed accretion disks we consider disk models for different values of free
parameters such as binary separation $a$ and  mass loss rate $\dot M_\mathrm{w}$. 
Binary and planet orbits are circular ($e=0$). For each model we set $v_\mathrm{w}$ = 20 \kms, $ c_\mathrm{w}$ = 10 \kms, $M_1$ = 5 M$_{\odot}$, $M_2$ = M$_{\odot}$, $R_2$ = R$_{\odot}$, $\alpha$ = 0.01.
 The total rate of mass capture (\ref{eq:mdot_acc}) and the Bondi radius eq. (\ref{eq:R_a}) are obtained for each set of parameters $a, \dot{M_\mathrm{w}}$. This specifies profiles of $\Sigma(r)$ and $H(r)$ for each disk model. Further, we solve differential equations (\ref{eq: type_I_speed}) and (\ref{eq: type_II}), respectively, for type I and type II migration in order to obtain the full time it takes the planet to reach the host star. Below we denote the latter by $t_\mathrm{m}$.

In Sec. \ref{section_The_SD_and_DMS_disk_profiles} and \ref{Migration in the SD and DMS disks} it is assumed that the primary
loses mass at a constant rate $\dot{M}_\mathrm{w}$, in Sec. \ref{Section_Migration_DSM__time-dependent_wind} the mass loss rate varies over time.

\begin{figure}
\centering
\includegraphics[scale=0.8]{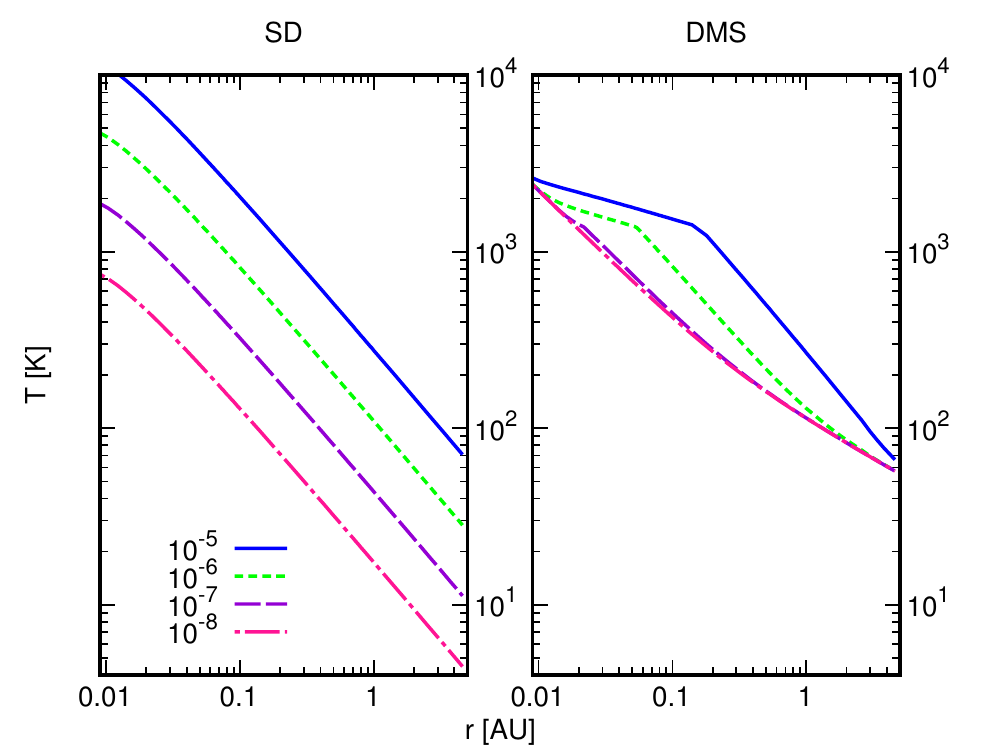}
\caption{\label{fig:Disk-profiles2} Midplane temperature profiles: the left and right panels correspond to the SD and DMS models, respectively. Line styles refer to different mass loss rates of the red giant (shown in the legend in solar masses per year), binary separation is set to a constant value 30 AU.}
\includegraphics[scale=0.8]{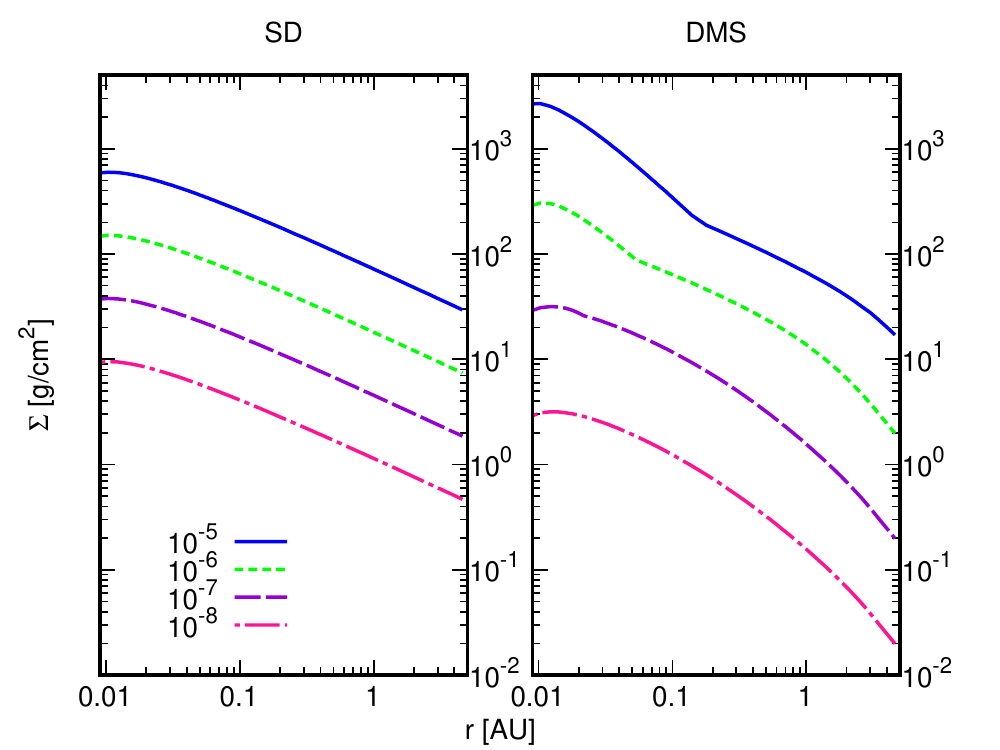}
\caption{\label{fig:Disk-profiles1} Surface density profiles: the left and right panels represent the SD and DMS models, respectively. Line styles refer to different mass loss rates of the red giant, binary separation is set to a constant value 30 AU. }
\end{figure}

\begin{figure}
\includegraphics[scale=0.8]{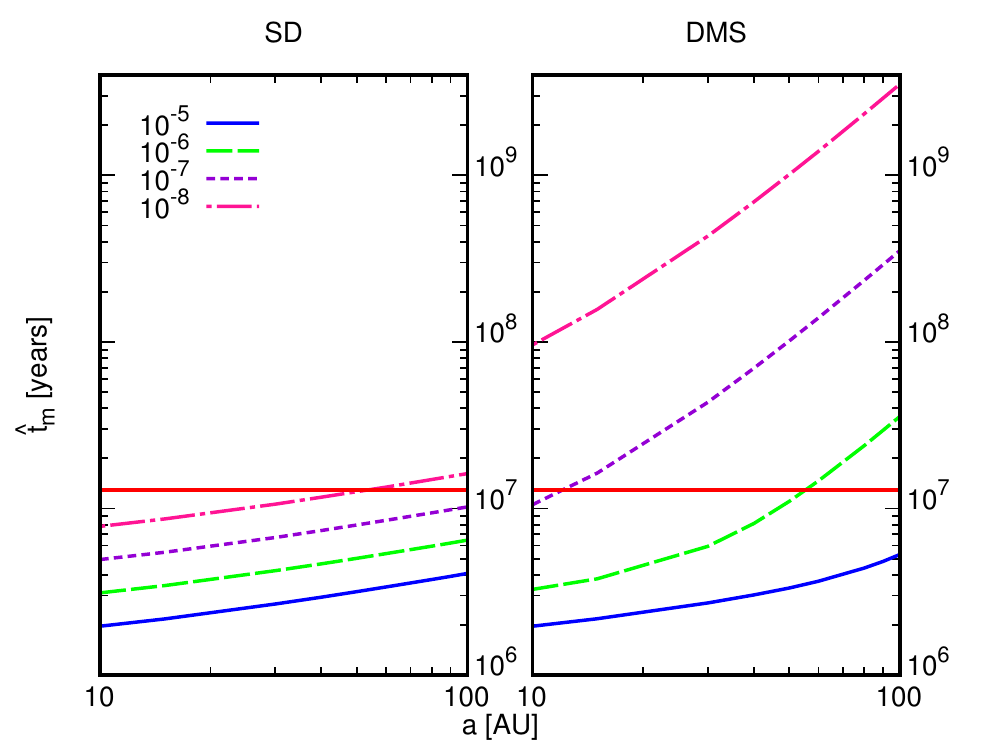}
\includegraphics[scale=0.8]{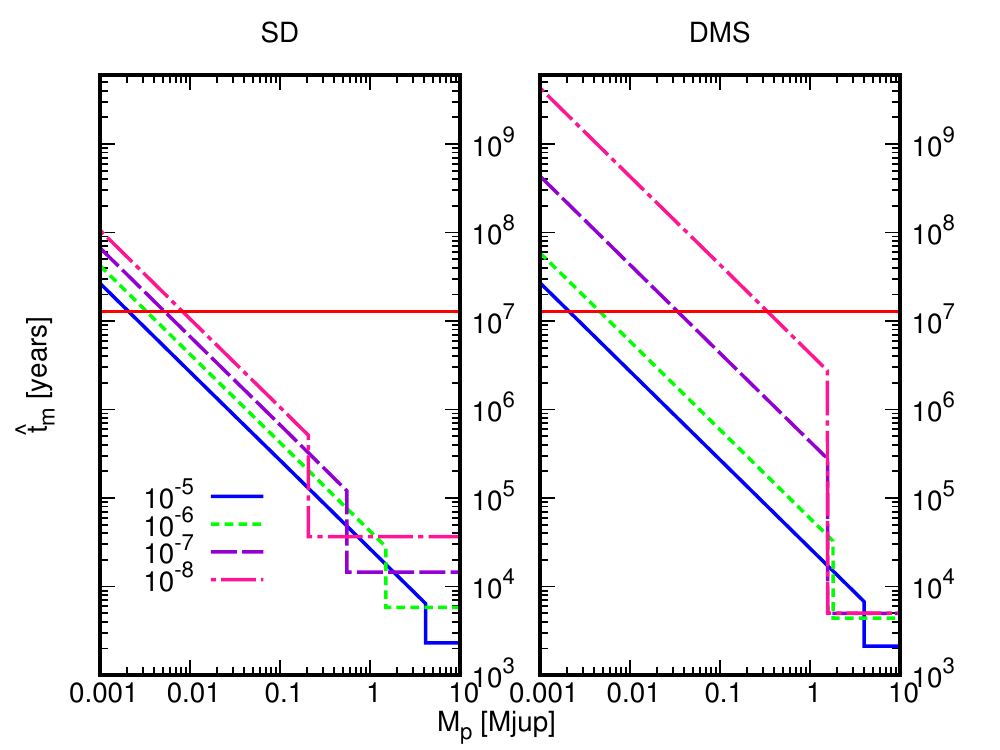}
\caption{\label{fig:a1}
Migration time scale $\hat t_\mathrm{m}$ for a planet at the orbital distance 1 AU.
Top panel: type I migration time dependence on binary separation for 0.01 M$_\mathrm{Jup}$ planet.
Bottom panel: types I, II migration time dependence on planet mass. Binary separation is set to 30 AU. Line styles refer to the different mass loss rates of red giant (shown in the legend in solar masses per year). Left and right panels are obtained for SD and DMS models, respectively. Horizontal thick line refers to the disk lifetime which is limited to the duration of red giant phase of the donor with mass 5 M$_{\odot}$ \citep{padova}. Calculations are made for the constant mass loss rate.
}
\end{figure}

\begin{figure}
\includegraphics[scale=0.8]{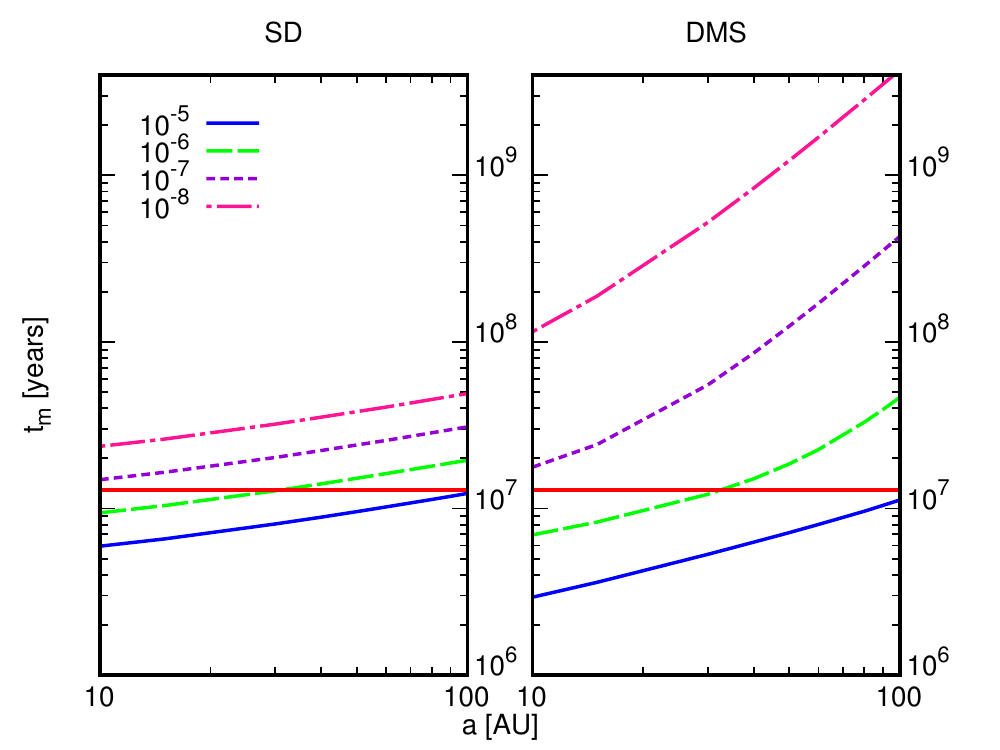}
\includegraphics[scale=0.8]{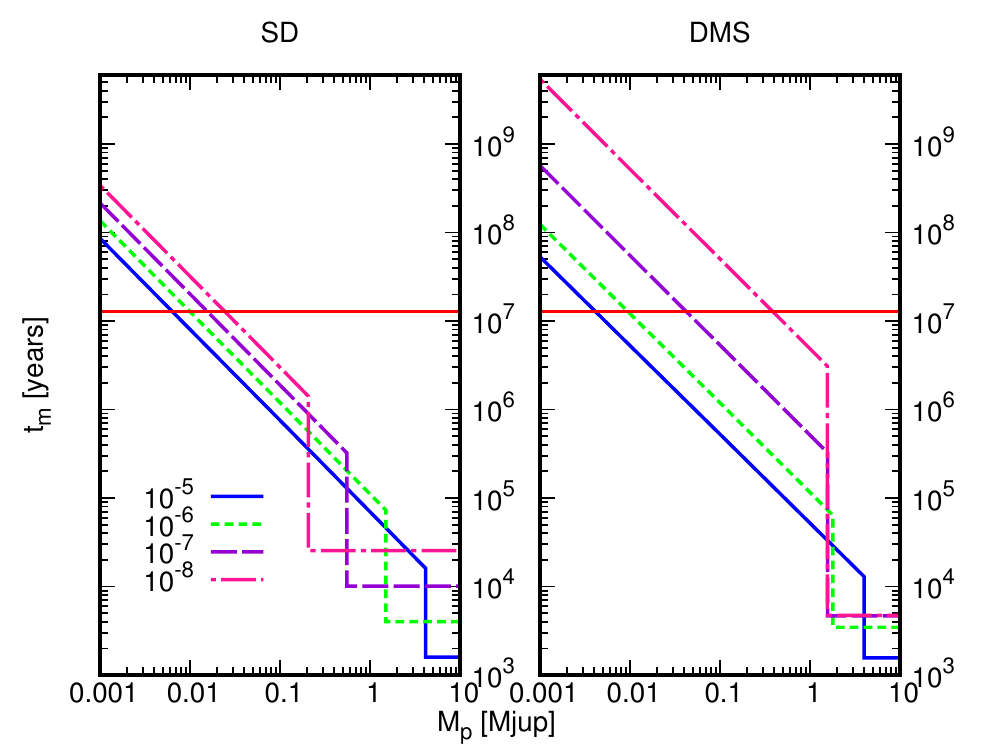}
\caption{\label{fig:a1_integr}
The full migration time, $t_\mathrm{m}$, for the same free parameters as in Fig. \ref{fig:a1}.
}
\end{figure}

\subsection{SD and DMS profiles} \label{section_The_SD_and_DMS_disk_profiles}

For representative binary separation $a=30$ AU and mass loss rate in the range $\dot{M}_{w}\sim10^{-8}-10^{-5}$ \msunyr\  Eq. (\ref{eq:mdot_acc}) yields the accretion rates $\dot{M}^{\rm tot}_{\rm acc} \sim 2.5\times( 10^{-11}-10^{-8})$ \msunyr\  which are typical for mature protoplanetary disks. Thus, we expect that wind-fed accretion disks resemble evolved low-mass protoplanetary disks, see also \citet{perken13}.
In this Section we compare the disc midplane temperature and surface density profiles for DMS model with that of SD model. These profiles are shown in Figs.~\ref{fig:Disk-profiles2},~\ref{fig:Disk-profiles1}. 

First, as soon as the disc is fed by strong stellar wind with the rate higher than $\sim 10^{-6}$ \msunyr\, its structure is determined predominantly by viscous heating (note, that such a strong wind can exist just for a relatively short period of time). In this case, both models give quantitatively similar distributions of $\Sigma$ and $T_\mathrm{c}$ at the sufficiently large radii. Thus, the disc is insensitive to modification of angular momentum balance introduced by the settling wind material, cf. eqs. (\ref{nu_sigma_SD}) and (\ref{nu_sigma}) as well as the subsequent eqs. (\ref{f_SD}) and (\ref{f}). 
Additionally, the contribution of the falling matter to the disk heating (see eq. \ref{eq:F_accr}) is small compared to 
viscous dissipation, 
which can be shown with the help of eqs. (\ref{nu_sigma_SD}), (\ref{f_r_in<<r<<r_a}) and (\ref{eq: t-visc1}): 
\begin{equation}
\label{T_fall_T_nu}
\frac{T_{\rm fall}^4}{T_{\nu}^4} \sim \frac{r}{r_\mathrm{a} \tau}
\end{equation}
provided that the heating rate 
due to the falling matter 
is given by eq. (\ref{eq:F_accr}) and one considers the disk energy balance far from both $r_\mathrm{in}$ and $r_\mathrm{a}$.

The change of the slope of the DMS temperature profile at small radii $\lesssim 0.2$ AU 
indicates significant opacity modification according to eq. (\ref{kappa}) as the evaporation of dust is included into this model. 
Drastic decrease of the opacity leads to much lower midplane temperature in the region of dust evaporation in contrast to the SD case. 
Since the surface density is inversely proportional to the temperature for the same value of the accretion rate, see eq. (\ref{Sigma_T}), this region 
corresponds to the increased surface density of DSM as compared to SD (see Fig.~\ref{fig:Disk-profiles1}). Thereby, in the case of strong stellar wind  inner parts of DMS are generally cooler and denser than its SD analogues. 



Disks fed by a weak stellar wind with the rate lower than $\sim 10^{-7}$ \msunyr\ are strongly affected by stellar irradiation. In contrast to SD which becomes progressively cooler as the accretion rate decreases, the temperature profiles of the corresponding DMS virtually coincide with each other as seen on the right panel in Fig.~\ref{fig:Disk-profiles2}. 
According to eq. (\ref{Sigma_T}), 
the surface density of such DMS is decreased in comparison with the corresponding SD obtained for the stellar wind of the same intensity. 
Thereby, in the case of weak stellar wind DMS is generally hotter and less dense than SD at
all distances from the host star (see two bottom lines on left and right parts of Fig. \ref{fig:Disk-profiles1}). 
At the same time, DMS are not hot enough to evaporate dust and break the profiles in this case. 

\subsection{Migration in the SD and DMS}
\label{Migration in the SD and DMS disks}
At first, we estimate the migration time scale at $r=1$ AU, see Fig.~(\ref{fig:a1}): 
\begin{equation}
\hat t_\mathrm{m}=\dfrac{r_\mathrm{p}}{dr_\mathrm{p}/dt}, \label{evaluation}
\end{equation}
where $dr_\mathrm{p}/dt$ is defined by eq. (\ref{eq: type_I_speed}) for type I  and eq. (\ref{eq: type_II}) for type II migration, correspondingly.  
Generally, type I migration time scale (see the top panels in Fig. \ref{fig:a1}) grows either with a binary separation or with a decrease of the mass loss rate of the donor in both SD and DMS models, since in the both cases the host star captures less wind material constructing an accretion disk of smaller density. Furthermore, in the case of DMS this tendency looks much stronger. Particularly, in the case of DMS $\hat t_\mathrm{m}$ exceeds the disc lifetime for $\dot M_\mathrm{w} \lesssim 10^{-8} ~\msunyr$
and $M_1\sim$ 5 M$_{\odot}$  throughout the considered range of $a$ in sharp contrast to that of SD. Also, for the particular $\dot M_\mathrm{w}$ migration slows down faster in the case of DMS as one proceeds to binaries with larger separations.   
The reason for such difference between migration rates in SD and DMS is caused mostly by stellar irradiation, which makes DMS less dense for the same $\dot M_\mathrm{w}$, see Section \ref{section_The_SD_and_DMS_disk_profiles}.



The transition from type I to type II migration, see the bottom panels in Fig. \ref{fig:a1},  occurs when the planet-to-star mass ratio starts to exceed the critical value defined by eq. (\ref{eq: q_crit}). 
This ratio 
depends on the disk vertical scale height and, accordingly, on both the mass loss rate of the red giant and the binary separation. Hence, for different curves in Fig. \ref{fig:a1} transition to type II migration occurs for different planet mass. Generally, the higher is the mass loss rate of the red giant, the faster are both type I and type II migrations and the higher planet mass corresponds to the transition between them. 
However, the heating of disk by stellar irradiation included in DMS speeds up type II migration and defers the transition from type I migration as one approaches the Jovian mass. 
Combination of the slowdown of type I migration with the speedup of type II migration in this case leads to an abrupt increase of the migration rate as gap is opened around the planet.


Note that for disk models considered here type I migration always proves to be slower 
than type II migration. Thus, we do not find rapid type I migration revealed in the studies of protoplanetary disks (see e.g. \citet{Bate}), when $\hat t_\mathrm{m}$ would be shorter than the corresponding type II migration time scale.
The model used by 
\citet{Bate} is constructed for fiducial constant aspect ratio $H/r=0.05$, which 
provides a denser disk as compared to 
SD with the same $H/r$. For example, in \citet{Bate} $\Sigma=50$ g cm$^{-2}$ at 5 AU, whereas for SD $\Sigma=10$ g cm$^{-2}$ at the same distance. We check that both SD and DMS acquire a higher surface density as one changes to the opacity of icy dust, $\kappa \propto T_\mathrm{c}^2$, appropriate in a disk beyond the snowline resulting in the increase of type I migration rate. However, in this study we consider migration inside the snowline.



To make sure that $\hat t_\mathrm{m}$ provides a good estimate of time that planet spends to get close to the host star, we obtain full migration time $t_\mathrm{m}$, see Fig. \ref{fig:a1_integr}. The value $t_\mathrm{m}$ is obtained integrating Eq. (\ref{eq: type_I_speed}) and Eq. (\ref{eq: type_II}), respectively, for type I and type II migration. The integration is terminated as planet attains distance, where tidal dissipation effects become strong enough to keep on the approach of the planet to the star surface. This final distance is estimated using Eq. (\ref{r_fin}). 
As can be seen in Fig. \ref{fig:a1_integr}, $t_\mathrm{m}$ is higher than $\hat t_\mathrm{m}$
by a factor of few. Moreover, this correction is less noticeable in the case of DMS.

 


\subsection{Migration in the DMS fed by time-dependent stellar wind}\label{Section_Migration_DSM__time-dependent_wind}

In this section we present our main results.
We consider migration in DMS with the account of time-dependent stellar wind. 
To obtain the instant $\dot{M}_\mathrm{w}$  we use Eq. (\ref{eq:dotMw}) along with Padova database of stellar evolutionary tracks \citep{padova} which provides us with $L_1$ and $R_1$ for each moment of time. 
Differential equations  (\ref{eq: type_I_speed}) and (\ref{eq: type_II}) are solved, respectively, for type I and type II migration including the variations of $\Sigma$ and $H$ due to variations of the mass loss rate of the red giant.

\begin{figure}
\includegraphics[scale=0.8]{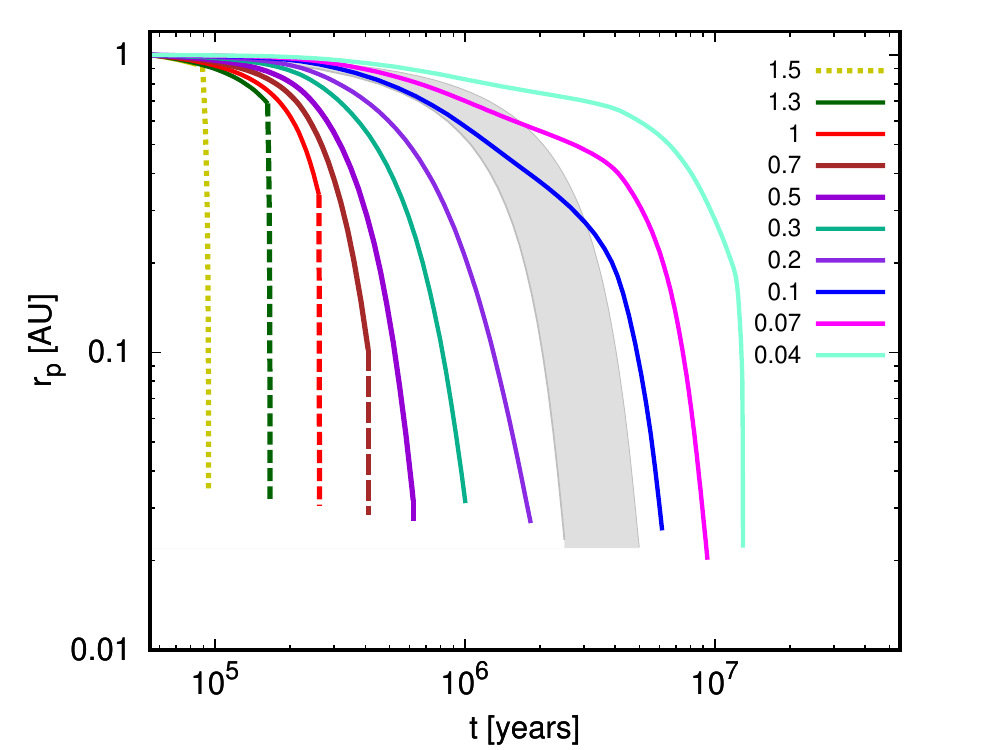}
\caption{\label{fig:a_t} Evolution of semi-major axes of planets migrating in DMS fed by stellar wind alternating over time. The dashed and the solid parts of each curve represent, respectively, type I and type II migration. Different styles of dashed curves introduce the planet mass given in units of the Jupiter mass (see the legend). Planet masses monotonically decrease from left (lower) curves to right (upper) curves. Grey region refers to migration of planets with masses 0.1-0.2 M$_\mathrm{Jup}$ in disk fed by constant wind with mass loss rate $10^{-8}$ \msunyr . Note, that for low masses the transition to type II migration does not occur. Binary separation is set to 10 AU and donor star mass is 5 M$_{\odot}$. }
\end{figure}

\begin{figure}
\includegraphics[scale=0.8]{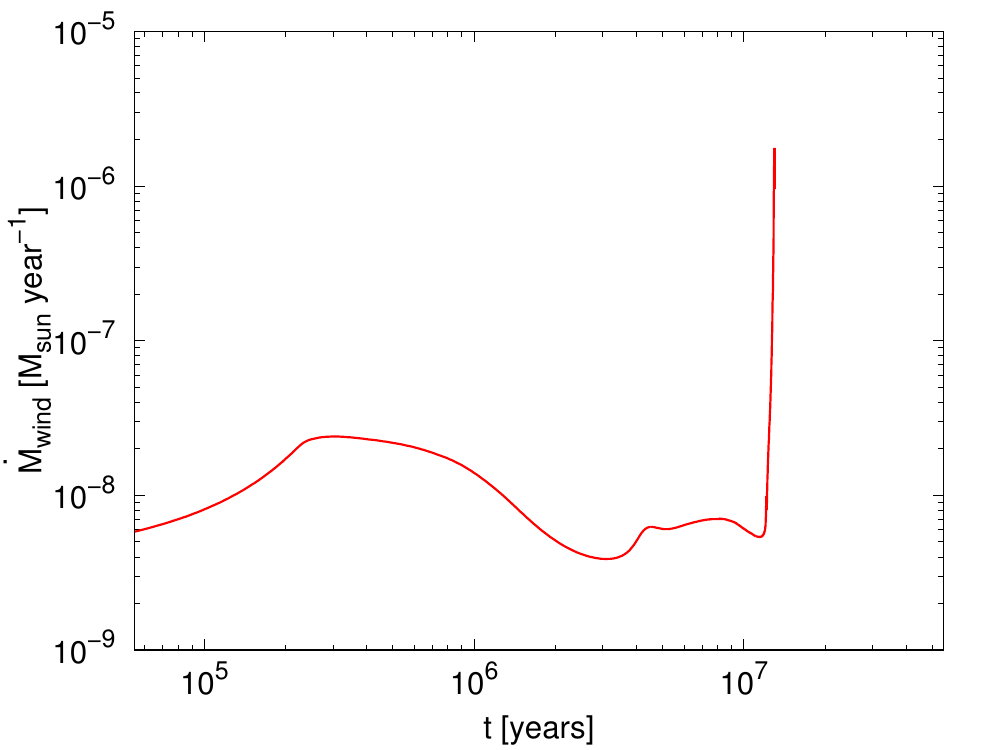}
\caption{\label{fig:dotM} Mass loss rate of the giant changing with time. The giant mass is 5 M$_{\odot}$. Mass loss rate is calculated using eq. 
(\ref{eq:dotMw}) and Padova stellar evolutionary tracks \citep{padova}.}
\label{wind}
\end{figure}

Evolution of semi-major axes of planets in this situation is shown in Fig. \ref{fig:a_t} for the particular choice of donor mass and binary separation.
We conclude that even strong variations of stellar wind during the red giant phase of the primary
leave the basic result unchanged, i.e. planet has enough time to approach the star surface in wide range of masses.
The shape of the curves in Fig. \ref{fig:a_t} that represent migration of planets with mass less than 0.2 M$_\mathrm{Jup}$ reflects the behaviour of stellar wind over time.
One finds that migration slows down from $t\gtrsim 10^6$ yr, accelerating again 
from $t\gtrsim 4\times 10^6$ yr. Additionally, the final cutoff seen for curve corresponding to 
0.04 M$_\mathrm{Jup}$ is related to the dramatic increase of the donor mass loss rate at the end of the red giant phase (see Fig. \ref{wind}). 

The transition from type I to type II migration occurs at various distance from the host star depending on the planet mass. For more massive planets it occurs at larger distances, while it is absent for sufficiently low masses such as $0.01-0.2$ M$_\mathrm{Jup}$. As soon as migration is changed to type II, planet rapidly moves towards the star.


\section{Discussion}

In this section we briefly discuss a number of simplifying assumptions of our model that can be relaxed in the future work.


\subsection{Role of irradiation by the companion}

The size of the red giant changes dramatically with time. At some point, the stellar radius becomes large enough to irradiate the surface of the disk, which may significantly increase its temperature since the red giant has large luminosity. This effect was not taken into account in our model. However, the case when it is important can be identified by the following condition: 

\begin{equation}
\beta_\mathrm{disk}=\beta_\mathrm{giant} ,
\label{eq: angles1}
\end{equation}
where $\beta_\mathrm{disk}$ is the disk grazing angle and $\beta_\mathrm{giant}$ is an angle between the orbital plane and the direction from the disk inner edge to the red giant tangent to its surface. In this study we consider disks existing from the beginning  of the RG stage till condition (\ref{eq: angles1}) is reached. For the 5 M$_{\odot}$ donor Eq. (\ref{eq: angles1}) is satisfied after $1.25 \times 10^{7}$ years after the beginning of the RG phase, which is close to the lifetime of the RG.

\subsection{Improved model of migration}


Numerical studies of planet migration in non-isothermal protoplanetary disks which take into account thermal and viscous diffusion show that type I migration of super-Earths substantially decelerates or even reverses outwards, see \citet{Paardekooper_Mellema_2006}, \citet{Kley_Crida_2008} and \citet{Kley_Klahr_2009}.
Analytic considerations, e.g. \citet{Baruteau_Masset_2008} and \citet{Paardekooper_Papaloizou_2008}, reveal a key role of U-turn motions of gas in the horseshoe region around a planet, which contribute to corotation torque. The latter remains unsaturated in the course of planet migration in the case the time of viscous and thermal diffusion in a disk are comparable to characteristic period of U-turn motions.
The corresponding modification of type I migration should have implications for planet orbital evolution in wind-fed accretion disks considered in this paper. Analytic formulae for migration torque with the account of dissipative effects suggested by \citet{Paardekooper2010}, \citet{Paardekooper2011} along with \citet{Masset_Casoli_2010} and \citet{Jimenez_Masset_2017} can be used for more consistent treatment of type I migration in such disks. 



Another effect which should be important for low-mass wind-fed disks is  slowdown of the type II migration due to  inertia of a planet when matter in the  disk  piles up beyond the gap. That is, if the planet becomes more massive than the part of the disk inside the planet's orbit, $M_\mathrm{d}$, the migration rate is reduced by the factor $\sim M_\mathrm{p}/M_\mathrm{d}$ according to \citet{Ivanov_1999}. At the same time, the migration timescale should be less than the characteristic time of accretion onto the planet which is $M_\mathrm{p}/\dot M^{\rm tot}_{\rm acc}$ in our notations. For the least mass loss rate of the donor considered here $\dot M_w\sim 10^{-8} ~\msunyr$\ the latter becomes $\sim 10^8$ yr  for planet of Jovian mass, which is significantly longer than the duration of the RG phase for donor with mass 5 M$_\odot$.

\subsection{Orbital eccentricity and inclination}



In this study we consider only circular orbits under the assumption that planet orbit lies in the disk plane. However, eccentric orbits, as well as orbits inclined with respect to the host star rotation axis and with respect to each other are common in planetary systems \citep{Winn_Fabrycky_2015}.

Rotation axes of binary components 
may be inclined with respect to the orbital plane of the binary \citep{1994AJ....107..306H}, and planets can be formed in such systems \citep{2018ApJ...861..116Z}. Inclination and shape of an s-type planetary orbit can evolve in a binary system due to Lidov-Kozai effect \citep{lidov, kozai}.
In addition, the orbital plane can be modified in a system where the host star  spin axis and orbital angular momentum of the binary are not aligned \citep{2012Natur.491..418B}. This suggests that generally, at least in the beginning of the red giant phase of the primary, the orbit of a pre-existing planet is not necessarily circular and/or is not aligned with the plane of a newly formed wind-fed disk (however, recent studies provide arguments that misalignment of stellar spin axes with the orbital angular momentum is more an exclusion than a typical situation, see \citealt{2018MNRAS.478.1942S}). How this affects planet migration in wind-fed accretion disks is a subject of future research. Eccentricity and inclination of planet orbits are known to be damped by interaction with protoplanetary disk. For values of those quantities less than the disk aspect ratio $\sim H/r$ damping rate is higher than the migration rate by a factor $\sim (r/H)^2$, --- see the linear analysis by \citet{Tanaka_Ward_2004}, --- thus, weakly affecting the migration. In contrast, for values of inclination and eccentricity larger than $\sim H/r$ the damping action of the disk is strongly reduced. The corresponding estimates of inclination decay rate in the dynamical friction approach give the characteristic time somewhat less than the migration rate, see \citet{Rein_2012} and \citet{Papaloizou_2013}. Numerical simulations performed for massive planets $M_\mathrm{p} > 1$ M$_\mathrm{Jup}$ on highly inclined orbits confirm that the inclination decay rate decreases for increasing initial inclination and increases for increasing planet mass. At the same time, the migration rate is considerably reduced at high inclinations \citep{Papaloizou_2013}.




\subsection{Migration of distant planets}


We do not address migration of planets from the outer parts of the disk beyond the snow line. However, mostly planets (and protoplanets) are formed in the zone of ice condensation beyond the snowline of protoplanetary disk, which normally corresponds to distances from several to ten AU for different host stars and molecules. 
Formation of a new disk, as suggested here, should initiate  migration of these distant bodies closer to the secondary star, possibly followed by a subsequent multiple planet-star mergers. 
Such a problem is straightforward to solve incorporating the transition to the opacity of icy dust in the disk model. Similar to the situation in protoplanetary disk, see e.g. \citet{Bell_1997}, wind-fed disks are expected to be relatively cold and dense beyond its own snowline, providing a faster type I migration than considered above.



\subsection{Binaries with less massive donors}

In the examples above the mass of the donor was assumed to be 5 M$_\odot$. Such stars have mass loss rate at the RG stage $\gtrsim 10^{-8}$ M$_\odot$~yr$^{-1}$ (see Fig. \ref{wind}).
Our estimates show that less massive donors with mass loss rate below $10^{-8}$ M$_\odot$~yr$^{-1}$ give birth to disks, which are optically thin, at least, for $r_\mathrm{p}\gtrsim 1\,$AU. According to stellar evolutionary tracks, a star with a mass less than 5 M$ _\odot$ remains at the RG phase for much longer time up to $\sim 10^8$ yr, however, they have substantially weaker winds. This implies that planet migration in optically thin wind-fed disks should take place in binaries with less massive donors, which are more common. In this study we considered only optically thick disks, however, the $\alpha-$disk model can be generalised to the optically thin case, see e.g. \citet{Artemova_2006}.

\subsection{Additional growth of planets}

On average wind-fed disks must be less dense than young protoplanetary disks (i.e.  disks at the start of planet formation) questioning whether the new planets may grow via the standard ``bottom-up'' scenario of dust coagulation and planetesimal accretion. Nevertheless, this may be compensated by the long existence of such disks. Indeed, even for the smallest mass loss rate of the donor considered here $\dot M_\mathrm{w} \sim 10^{-8}$ \msunyr\ the total mass captured by disk along its lifetime $\sim 10^7$ yr approaches the Jovian mass. Further growth of pre-existing planets must be another important process to be studied in wind-fed disks in binaries --- both, for the analysis of new paths of planet formation and in connection with planet-star mergers since an additional planet growth enhances planet migration. It would be interesting to consider the particular case of pre-existing super-Earths and icy giants in this context, as they represent probably the largest population of planets.   

The core accretion formation of giant planets adapted for the case of wind-fed disks in binaries suggests that the outcome of an additional planet growth is controlled, at first, by the critical core mass for quasi-static accumulation of envelope and, at second, by the conditions in the surrounding disk which define the rate of the runaway growth. The critical core mass crucially depends on the rate of the envelope heating, which in turn is defined by the accretion rate of solid material onto the pre-existing planet. The amount of solid material either in the form of pre-existing debris, or in the form of newly coagulated peebles, or in the form of newly accumulated planetesimals is a unknown quantity. 
The issues just mentioned above deserve a detailed research, while the study of planet formation in ``next generation'' disks in binaries is at the beginning now \citep{Hogg_2018}.

\section{Conclusions}

In this paper we studied an idealized situation when a planet in a binary system is embedded into an accretion disk coplanar with its s-type orbit. The disk is formed by stellar wind of the second (more massive) evolved 5-solar mass companion. We analyse migration (I and II type) within the snow line for different types for two types of accretion: standard $\alpha$-disk and $\alpha$-disk with matter settling on the disk surface within the Bondi radius.   

We demonstrate that massive planets ($\gtrsim0.1$ M$ _\mathrm{Jup}$) can migrated in such a disk down to the stellar surface during the RG stage of the companion. Thus, binary systems with evolved companions can contribute to star-planet coalescence which can be potentially observed in the near future.  

\section*{Acknowledgements}

S.P. and V.Z. acknowledge the support from the Program of development of M.V. Lomonosov Moscow State University (Leading Scientific School 'Physics of stars, relativistic objects and galaxies').

    	\normalem 
    	\bibliographystyle{mnras}

\end{document}